\documentstyle[12pt,epsf,rotate]{article}
\input amssym.def
\input amssym.tex
\begin{document}
{\noindent \Large\bf Topology and Causal Structure}\\
\vspace*{0.2cm}

{\noindent \large \hspace*{2.5cm} Andrew Chamblin}\\
\vspace*{0.01cm}

{\noindent \small \hspace*{2.5cm} Department of Applied Mathematics and
Theoretical
Physics,}\\
{\noindent \small \hspace*{2.5cm} University of Cambridge, Cambridge CB3 9EW,
England}\\
\vspace*{0.4cm}

{\noindent \small \hspace*{2.5cm} {\bf Abstract.} We investigate the causal
structure
of spacetimes}\\
{\noindent \small \hspace*{2.5cm}  $(M, g)$ for which the metric $g$ is
singular on a set of points. In}\\
{\noindent \small \hspace*{2.5cm} particular, we show that it is always possible 
to continuously de-}\\
{\noindent \small \hspace*{2.5cm} form the metric so that the 
resulting spacetime is causal.}\\
{\indent \small \hspace*{2.5cm} Furthermore, we develop the framework of
$(2 + 1)$-dimensional}\\
{\noindent \small \hspace*{2.5cm} pure gravity to include
processes describing the creation and annih-}\\
{\noindent \small \hspace*{2.5cm} ilation of gravitational skyrmions, or
`kinks'. In this approach, a}\\
{\noindent \small \hspace*{2.5cm} sector in the `Fock space' is classified by
purely
topological data:}\\
{\noindent \small \hspace*{2.5cm} The topology of the triad and of the manifold
in
the asymptotic}\\
{\noindent \small \hspace*{2.5cm} region. We use this construction to
generalise
the selection rule of}\\
{\noindent \small \hspace*{2.5cm} Amano and Higuchi [5] to a
selection rule governing when the am-}\\
{\noindent \small \hspace*{2.5cm} plitude for a given process
vanishes. We present various applicat-}\\
{\noindent \small \hspace*{2.5cm} ions and examples of such processes.}\\
\vspace*{0.6cm}

{\noindent \bf 1.~ Metric Topology}\\

{Statements about the kinematical aspects of causal structure have long been of
interest to those working in classical and quantum gravity ([1], [2], [3],
[4]). For the
most part, these studies have focussed on the causal structure of spacetimes
$(M, g)$
for which the metric $g$ is globally non-singular. However, the work of a
number of
authors ([7], [8], [9], [10]) suggests that degenerate metrics may be a generic
feature of any quantum theory of gravity which incorporates topology changing
processes. With this in mind, it is natural to investigate the causal
properties of
spacetimes with degenerate Lorentzian structures. In order to fully appreciate
this
issue, we need to first define, in a detailed way, what a `singularity' in the
metric
is, and how the `index' of a singularity is associated with the topology of the
metric.

To begin with, let us assume that $(M, g)$ is $n$-dimensional and {\it
time-orientable} [11]. It follows [5] that we can express the Lorentz metric,
$g_{ab}^{L}$, in terms of some auxilliary Riemannian metric $g_{ab}^{R}$ and
some
timelike vector field $V$, in the usual way:}
\begin{equation}
{g_{ab}^{L} = g_{ab}^{R} - 2V_{a}V_{b}}
\end{equation}
{where in equation (1), $g_{ab}^{R}$ is some (arbitrary) Riemannian metric on
$M$, and
$V$ has been normalised with respect to $g_{ab}^{R}$, that is,
$g_{ab}^{R}V^{a}V^{b} =
1$. Thus, there is a correspondence between vector fields and Lorentz
metrics
on $M$.

Now, a natural question is: Given two Lorentz metrics, $g$ and $g^{\prime}$ (on
$M$),
when can $g^{\prime}$ be obtained from $g$ by a continuous deformation?
This question should be relevant in any Lorentzian sum-over-histories
version of quantum gravity, since if two metrics $g$ and $g^{\prime}$ are
related
by some element ${\phi}$ of the identity connected component of the
diffeomorphism group it follows that the two metrics must also be
homotopic.  In the light of
the above correspondence between timelike vector fields and Lorentz metrics, we
see
that this question reduces to the question: Given two vector fields, $V$ and
$V^{\prime}$ (on $M$), when can $V^{\prime}$ be obtained from $V$ by continuous
deformation? This question is well understood [6]; it turns out that the answer
depends upon the degrees of certain maps (defined by the vector fields $V$ and
$V^{\prime}$) from certain $(n - 1)$-dimensional submanifolds of $M$, into
$S^{n -
1}$. In
the context of General Relativity, the degree of such a map is often called the
`kink
number' [12]. We need to describe the kink number in more detail.

First, we let $S(M)$ denote the unit-sphere bundle over $M$. That is, at each
$p
{}~{\in}~ M$, the fibre, $S_{p}(M)$, of $S(M) ~{\longrightarrow}~ M$ is the `$n
-
1$-sphere
of directions' at $p$. Since ${\dim}(S_{p}(M)) = n - 1$ and ${\dim}(M) = n$, we
see
that the dimension of the bundle space $S(M)$ is $2n - 1$. Let ${\Sigma}$
denote any
connected, closed $(n - 1)$-dimensional submanifold of $M$. Then we can
restrict the
bundle $S(M) ~{\longrightarrow}~ M$ to ${\Sigma}$, obtaining the $2n - 2$
dimensional
bundle $S({\Sigma}) ~{\longrightarrow}~ {\Sigma}$, the $S^{n - 1}$-bundle over
${\Sigma}$. Since $(M, g)$ is time-orientable, we can naturally choose a unit
normal
vector field, $n$, to ${\Sigma}$. Clearly, $n$ determines a section of
$S({\Sigma}):
{}~n ~{\in}~ {\Gamma}(M, S({\Sigma}))$. Likewise, let $V$ be any future
timelike vector
associated with the Lorentz metric $g_{ab}^{L}$ (as in equation (1)), then the
restriction of $V$ to ${\Sigma}$ also yields a section $V ~{\in}~ {\Gamma}(M,
S({\Sigma}))$. These two sections, $n$ and $V$, will generically intersect at a
finite
number of points ${\chi}_{i} ~{\in}~ S({\Sigma}), ~i = 1, ... m$. We assign
each
intersection point a sign, ${\mbox{sign}}({\chi}_{i})$, defined by}
\vspace*{0.1cm}

\[
{{\mbox{sign}}({\chi}_{i}) = \left\{ \begin{array}{cl}
+1 &\hspace*{1cm} {\mbox{iff the orientation of $S({\Sigma})$ at ${\chi}_{i}$
equals
the product}} \\
 &\hspace*{1cm} {\mbox{of the orientations of $V$ and $n$ at ${\chi}_{i}$}} \\
 & \\
-1 &\hspace*{1cm} {\mbox{otherwise}}
\end{array} \right. }
\]
\vspace*{0.1cm}

{\noindent We can then define the kink number of the timelike vector field $V$
with respect to
the three-surface ${\Sigma}$ (denoted `${\mbox{kink}}({\Sigma}; V)$') by the
formula}
\begin{equation}
{{\mbox{kink}}({\Sigma}; V) = {\displaystyle\sum_{i =
1}^{m}}~{\mbox{sign}}({\chi}_{i})}
\end{equation}
{As discussed above, the topology of the Lorentz metric $g_{L}$ is equivalent
to the
topology of the timelike vector field $V$, which in terms of the kink number
simply
means that}
\begin{equation}
{{\mbox{kink}}({\Sigma}; V) = {\mbox{kink}}({\Sigma}; g_{L})}
\end{equation}
{With this definition of metric homotopy, we can now discuss the topology of
metric
singularities.

In this paper, we shall only consider the situation where 
the metric $g_{L}$ fails to be invertible on a collection
of points, $p_{i} ~{\in}~
M$; we shall refer to such a point where the metric is
not invertible as a {\it singularity}.  We should point out that
this sort of singularity is highly non-generic, since in 
general the metric will fail to be invertible on higher
dimensional hypersurfaces in the spacetime.  We adopt this
definition of a singularity purely as a simplifying assumption.
Using
equation (1), we see that a metric singularity is thus
present at any point $p ~{\in}~ M$ at
which
the timelike vector field, $V$, becomes singular, i.e., $V$ must vanish
identically at
$p$. The topological nature of vector field singularities is well-understood
[6]. To
each singularity, $p$, of a vector field $V$ on $M$, one assigns an {\it
index},
defined as follows:

Let $S^{n - 1}_{\varepsilon}(p)$ be a tiny $(n - 1)$-dimensional sphere (of
radius
${\varepsilon}$) at $p$. Since the vector field $V$ is assumed to vanish only
at $p$
(and not on a neighbourhood of $p$), $V$ is non-vanishing on $S^{n -
1}_{\varepsilon}(p)$. $V$ thus defines a map from $S^{n - 1}_{\varepsilon}(p)$
to $S^{n
- 1}$, and we define the {\it index of $V$ at $p$} (denoted ${\mbox{ind}}(V,
p)$) by the
formula}
\begin{equation}
{{\mbox{ind}}(V, p) = 1 - {\mbox{kink}}(S^{n - 1}_{\varepsilon}(p); V)}
\end{equation}
{In analogy with equation (3), we therefore define the index of a Lorentz
metric,
$g_{L}$, at a metric singularity $p$ by the relation}
\begin{equation}
{{\mbox{ind}}(g_{L}, p) = 1 - {\mbox{kink}}(S^{n - 1}_{\varepsilon}(p); g_{L})}
\end{equation}

{We now illustrate the rough characteristics of different index singularities
as an
aid to the visual imagination.}
\vspace*{0.2cm}

{\noindent {\bf Index 1:} ~If the metric $g_{L}$ has index equal to 1 at some
point $p
{}~{\in}~ M$, it follows from equation (5) that ${\mbox{kink}}(S^{n -
1}_{\varepsilon}(p); g_{L}) = 0$, for any little $(n - 1)$-sphere about $p$.
Intuitively, this means that the light cones do not `spin around' as one
approaches
the singularity $p$. Below, we have drawn three typical exampes of index 1
singularities, which we call `bang', `crunch', and `flush' singularities for
obvious
reasons: Fig. 1.}\\
\vspace*{0.5cm}

\epsfxsize=12cm
\epsfysize=15cm
\rotate[r]{\epsfbox{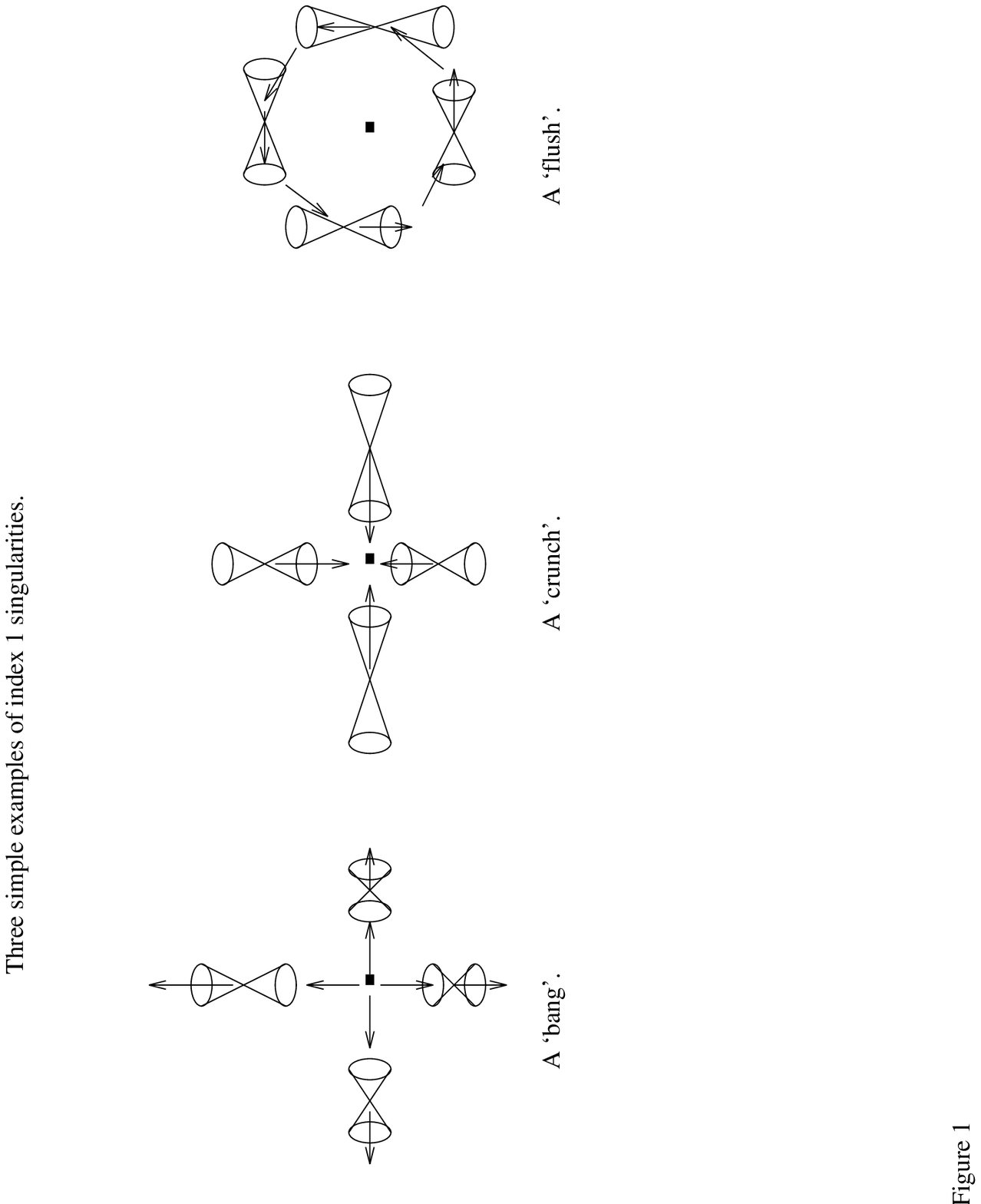}}
\vspace*{0.7cm}

{\noindent {\bf Index 2:} ~When the index of a metric singularity $p$ is 2, the
light
cone field must kink once `negatively' on $S^{n - 1}_{\varepsilon}(p)$, i.e.,}
\[
{{\mbox{kink}}(S^{n - 1}_{\varepsilon}(p); g_{L}) = -1}
\]
{A simple example of an index 2 singularity is the `dipole' singularity. In the
below
figure, we show how an index 2 singularity can be obtained from two index 1
singularities by continuously deforming the two index 1 singularities until
they `hit'
each other: Fig. 2.}\\
\vspace*{0.5cm}

\epsfxsize=12cm
\epsfysize=15cm
\rotate[r]{\epsfbox{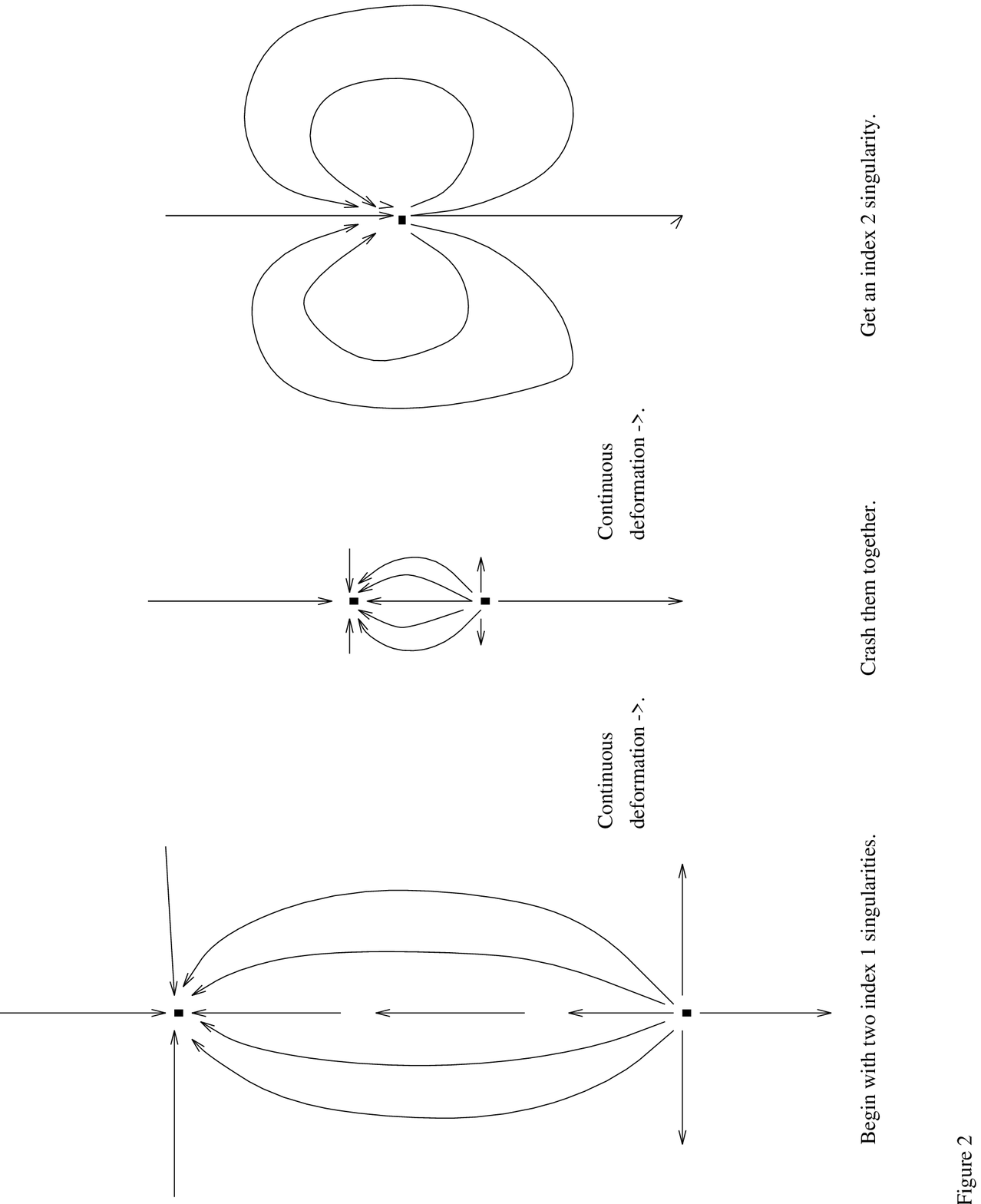}}
\vspace*{0.7cm}

{\noindent {\bf Index ~-1:} ~A metric $g_{L}$ with an index $-1$ singularity
$p$, must
have ${\mbox{kink}}(S^{n - 1}_{\varepsilon}( \\
p); g_{L}) = 2$. Interestingly, this is
precisely the type of singularity which occurs in the Deutsch-Politzer time
machine in
even spacetime dimensions [14]. We therefore refer the reader to [14] for an
example
of this type of singularity. \\

Higher and higher index singularities can be readily constructed by
generalising the
constructions illustrated above (for example, by `throwing together' a bunch of
index
1 singularities). This can be an amusing exercise, for some people. The point
is,
there is no difference (up to a continuous deformation) between an index 1
singularity, and a pair of singularities, one with index 597 and the other with
index
-596 (since we can `crash' them together to get the index 1 singularity back).
On the
other hand, we do not really feel that index 597 singularities are `generic'.
In fact,
index 1 singularities seem almost `natural', and at least have the distinction
of
appearing in the cosmology literature where one talks of big `bangs' and
`crunches'
(although we have yet to hear of a `big flush').

On the other hand, there is a real sense in which the `crashing
together' of point singularities is not a physical process.
After all, a singularity should presumably be thought of
as a component of the boundary of the spacetime; pushing two
singularities together would therefore seem to alter the nature of the
boundary.  Since the boundary of a compact spacetime
contributing to a functional integral is usually assumed to comprise the
`initial' and `final' states (before and after `scattering')
then spacetimes with different boundaries must correspond to
completely different processes.  We should therefore outlaw
deformations which move singularities together.  Indeed, this
leads to an interesting way of thinking about these point
singularities:  Since a neighborhood of any singularity
is a portion of spacetime homeomorphic to
${S^{n - 1}} ~{\times}~ {\Bbb R}$, it makes sense to think of
a singularity as an `asymptotic state' corresponding to an
`incoming' (or `outgoing') $S^{n - 1}$ with a kink on it
(we will have more to say about kinky boundary conditions
later).  With this in mind, whenever we consider a `continuous
deformation of a metric' we shall always mean a deformation
which does not move any metric singularities together.}\\
\vspace*{0.6cm}

{\noindent \bf 2. ~Causality and Singularities}\\

{To begin with, we recall the precise statement of Geroch's result:}
\vspace*{0.2cm}

{\noindent {\bf Theorem.} (Geroch [1]) {\it Let $(M, g_{L})$ be a compact,
time-orientable spacetime with globally non-singular Lorentz metric $g_{L}$.
Suppose
that the boundary of $M$ is equal to the disjoint union of two compact
spacelike
three-manifolds, ${\partial}M ~{\cong}~ {\Sigma}_{1} ~{\cup}~ {\Sigma}_{2}$,
and that
there are no closed timelike curves in $(M, g_{L})$. Then ${\Sigma}_{1}$ and
${\Sigma}_{2}$ are diffeomorphic, and $M$ is topologically of the form
${\Sigma}_{1}
{}~{\times}~ [0, 1]$.}}
\vspace*{0.3cm}

{Thus, if ${\Sigma}_{1}$ and ${\Sigma}_{2}$ are not even {\it homeomorphic},
there
must exist CTCs in $(M, g_{L})$ (we shall assume time-orientability throughtout
this
paper). Of course, if we allow the metric $g_{L}$ to be singular, then the
above
result will not in general hold. As Geroch noted in his original proof of the
above Theorem, the timelike flow (induced by $g_{L}$) will induce a
diffeomorphism
from ${\Sigma}_{1}$ to ${\Sigma}_{2}$ {\it unless} the timelike curves close
back on
themselves {\it or} they hit singularities.

A natural question, then, is whether or not it is always possible to arrange
for all
the CTCs to hit singularities. Given a compact causality violating spacetime
$(M,
g_{L})$ (with or without boundary), such that $g_{L}$ is singular at some
finite
collection of points ${\{}p_{1}, p_{2}, ... p_{n}{\}}$, can we always `deform'
all of
the CTCs so that they `hit' the singularities? The answer is yes. In fact, by
slightly
modifying the arguments of [3], we can prove}
\vspace*{0.2cm}

{\noindent {\bf Theorem 1.} {\it Let $M$ be a smooth compact $n$-manifold,
without
boundary, and let $g_{L}$ be some Lorentz metric on $M$ such that $g_{L}$ is
singular
on a finite collection of points ${\{}p_{1}, p_{2}, ... p_{m}{\}}$ in $M$ and
$(M,
g_{L})$ is causality violating. Then there always exists a new metric,
$g_{L}^{\prime}$, which can be obtained by a continous deformation of the
original
metric $g_{L}$ such that $(M, g_{L}^{\prime})$ is causal.}}\\
\vspace*{0.3cm}

{\noindent {\it Proof.} Given a point, $p_{i} ~{\in}~ M$, at which $g_{L}$ is
singular, let $B^{n}_{\varepsilon}(p_{i})$ denote the {\it closed} $n$-ball of
radius
${\varepsilon}$ about $p_{i}$ and $S_{\varepsilon}^{n - 1}(p_{i}) ~{\cong}~
{\partial}B_{\varepsilon}^{n}(p_{i})$, as above. Let $N$ be obtained from $M$
by
removing all of the $B_{\varepsilon}^{n}(p_{i})$:}
\[
{N ~{\cong}~ M - (B_{\varepsilon}^{n}(p_{1}) ~{\cup}~
B_{\varepsilon}^{n}(p_{2}) ...
{}~{\cup}~ B_{\varepsilon}^{n}(p_{m}))}
\]
{Then, by construction, the spacetime $(N, g_{L})$ is non-singular and {\it
open}
(since the boundary, ${\partial}N$, does not lie in $N$; this is the main
technical
difference from the construction in [3]). We wish to `cut' all of the CTCs in
$(N,
g_{L})$ by deforming them onto ${\partial}N$, as outlined in [3]. Since
${\partial}N$
is not in $N$, this is a bit tricky. Technically, we need to briefly pass to
$({\bar
N}, g_{L})$, where ${\bar N}$ denotes the {\it closure} of $N$. $({\bar N},
g_{L})$ is
then a {\it compact} spacetime with boundary ${\partial}N$. As in [3], we can
cover
${\bar N}$ with a {\it finite} number of sets of the form}
\[
{S_{q_{i}} = {\{}x ~{\in}~ I^{+}(q_{i}) ~{\cap}~ I^{-}(q)~|~q ~{\in}~
I^{+}(q_{i}){\}}}
\]
{and we can take the sets in this cover to be locally causal (no CTC lies
entirely in
any one of the $S_{q_{i}}$). As in [3], we can use these sets to construct a
continuous
deformation of the metric $g_{L}$ on ${\bar N}$ to a new metric
$g_{L}^{\prime}$, such
that all of the CTCs are pushed onto the boundary, ${\partial}N$.
We now pass back to $(N, g_{L}^{\prime})$. We have thus pulled all of
the CTCs arbitrarily close to the singular points, as shown: Fig. 3.\\
\vspace*{0.5cm}

\epsfxsize=12cm
\epsfysize=15cm
\rotate[r]{\epsfbox{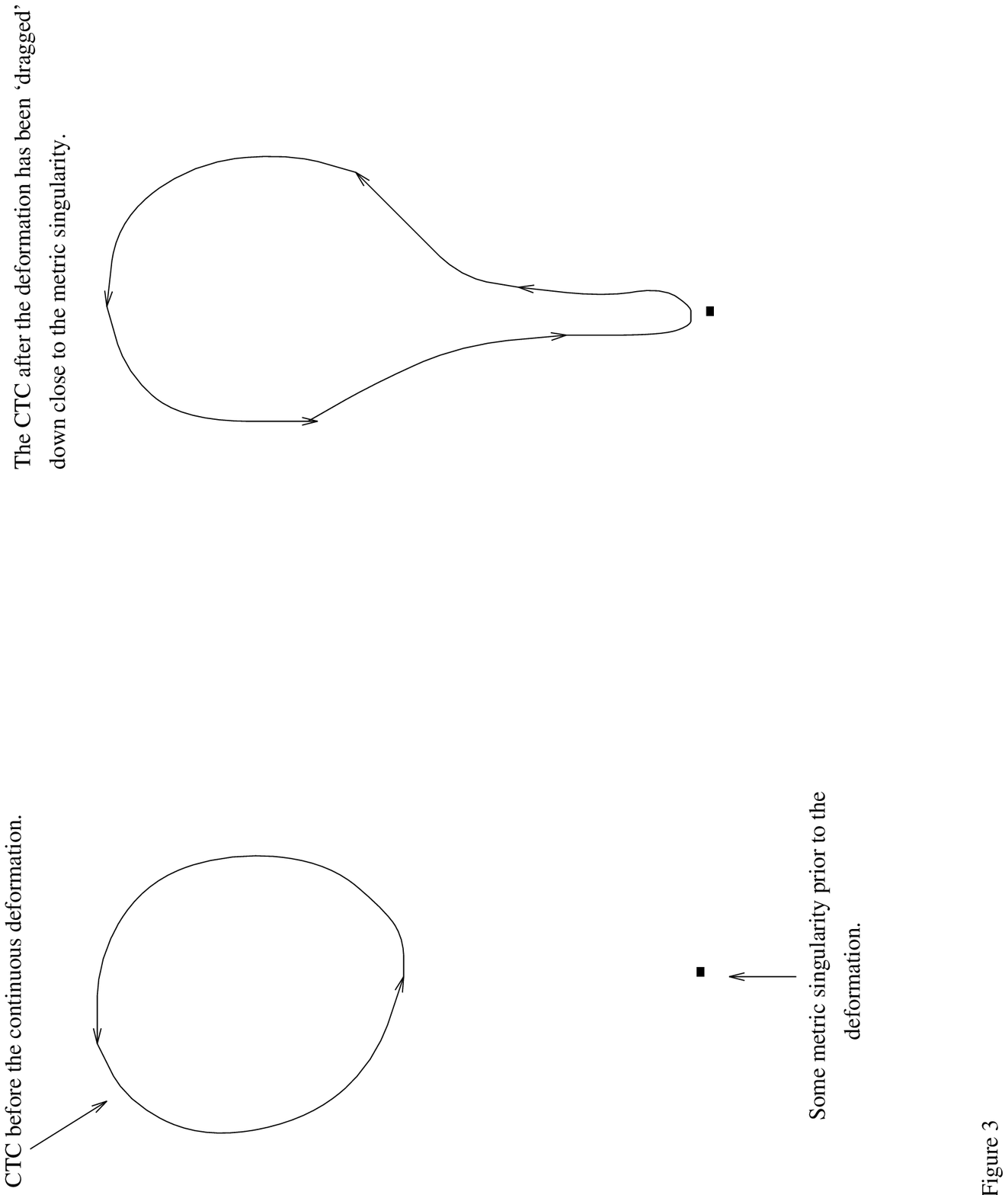}}
\vspace*{0.5cm}

Furthermore, since the deformation is continuous, the kink number on each
$S_{\varepsilon}^{n - 1}(p_{i})$ is unchanged, that is, we have not altered the
index
of any singularity. We then continue this construction in the limit
${\varepsilon}
{}~{\longrightarrow}~ 0$, so that all CTCs are continuously forced onto the
singular
points. We have therefore constructed a continuous deformation, $g_{L}
{}~{\longrightarrow}~ g_{L}^{\prime}$, such that $(M, g_{L}^{\prime})$ is
causal.
\hspace*{5cm} \hfill ${\square}$ \\
\vspace{0.2cm}

Thus, given a singular spacetime with CTCs, we can always `destroy' the CTCs
using the
singularities. Furthermore, by construction the deformation does not
move any of the singularities together.

Of course, one could take the view that {\it spacelike} topology change is too
restrictive and that the topology of the metric should also be allowed to
fluctuate.
In the context of a `Lorentzian' quantum gravity theory obtained by a path
integral
prescription, this would mean that instead of calculating the probability
amplitude to
pass from some initial {\it spacelike} three-manifold to some final {\it
spacelike}
three-manifold, one would instead
seek an `amplitude' describing the probability of transition `from' a
three-manifold
${\Sigma}_{1}$, such that the Lorentz metric $g_{L}$ has kink number $n_{1}$ on
${\Sigma}_{1}$, `to' a three-manifold ${\Sigma}_{2}$, such that the Lorentz
metric
$g_{L}$ has kink number  $n_{2}$ on ${\Sigma}_{2}$. In this way, the `boundary
data'
(or asymptotic states) of such a theory would be purely topological. We stress
this
viewpoint (even though it goes against the spirit of canonical quantum gravity)
since
we feel it is nonsensical to study strictly {\it spacelike} topology change via
spacetimes which
have metric singularities. To see why, let $(M, g_{L})$ be any spacetime with
spacelike boundary ${\partial}M$ and let $g_{L}$ be singular at $p$. We then
obtain a
spacetime $(M, g_{L}^{\prime})$, with a {\it kink} in $g_{L}^{\prime}$ on
${\partial}M$, simply be deforming the singular point $p$ {\it across}
${\partial}M$,
as shown: Fig. 4.\\
\vspace*{0.5cm}

\epsfxsize=12cm
\epsfysize=15cm
\rotate[r]{\epsfbox{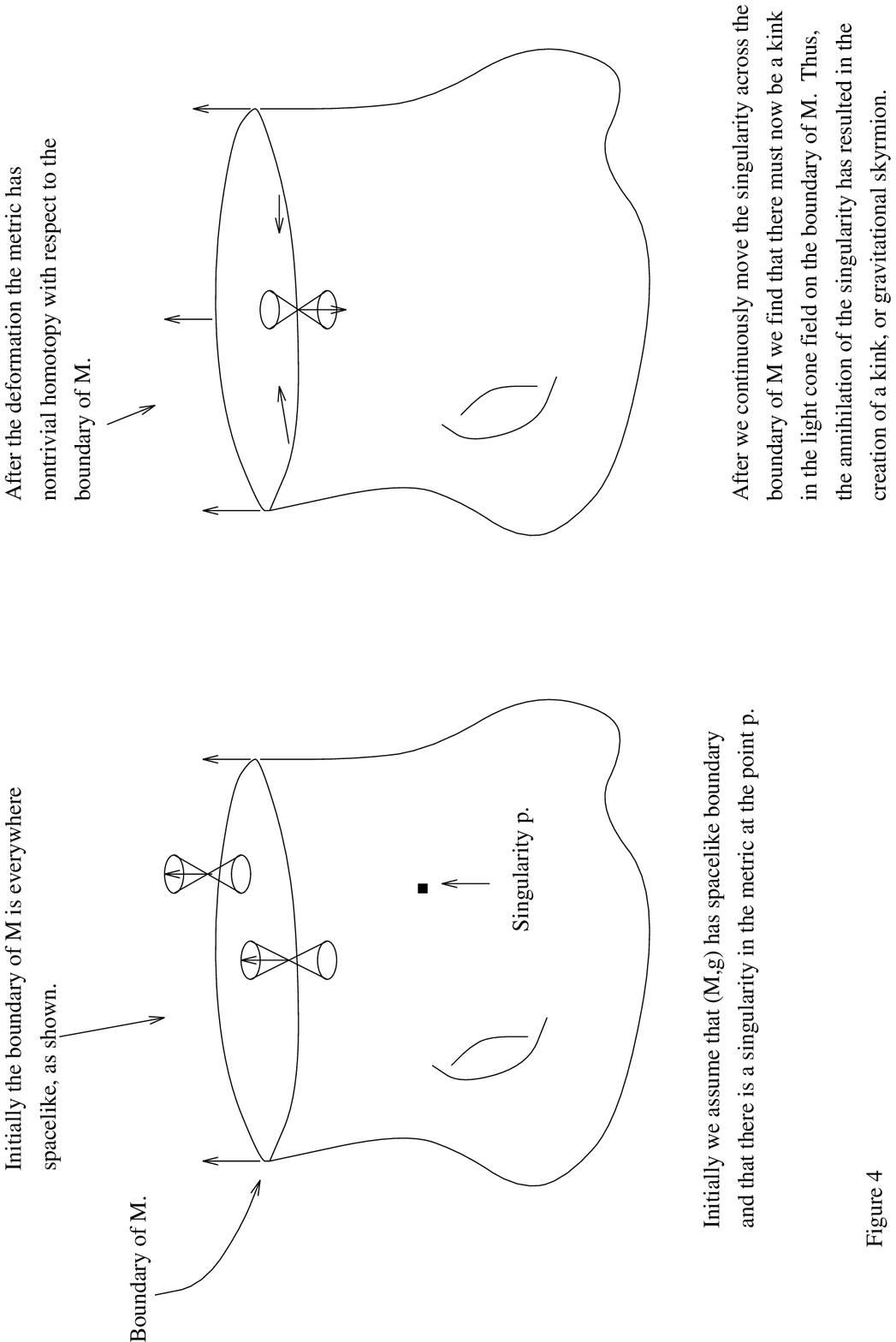}}
\vspace*{0.5cm}

The point is, the continuous deformation could perhaps
represent the
`flow of time'. The creation/annihilation of {\it singularities} corresponds to
creation/annihilation of {\it kinks} (note that in the above example, one would
have
${\mbox{kink}}({\partial}M; g_{L}^{\prime}) = {\mbox{ind}}(g_{L}, p)$). The two
concepts are dual in this sense (see also [18] for ideas along these lines).
This point of view also fits in nicely with our previous idea
that a singularity can (to a certain extent) be regarded as an
`internal' asymptotic state. We now present an example of a model where the
production and annihilation of kinks can be explicitly realised: $(2 \,+\,
1)$-dimensional gravity.}\\
\vspace*{0.6cm}

{\noindent \bf 3. The Model in $(2 + 1)$ Dimensions}\\

{Attempts to quantise gravity via a path integral or sum over histories
formalism have
led to many fascinating insights and results. One of the most interesting and
well-known
constructions to emerge from these efforts is Witten's $(2 + 1)$-dimensional
toy model
([10], [19], [20]). In this approach, one first writes down the theory in the
language of
fibre bundles. That is, one identifies, as the fundamental variables, a triad
$e^{a}$
and a spin ($SO(2, 1)$) connection $w^{a}$. The triad, which explicitly is just
a
collection of three linearly independent 1-forms, $e^{0}_{\mu}$, $e^{1}_{\mu}$
and
$e^{2}_{\mu}$, will in general be degenerate.
If we
recover the Lorentz metric by the usual formula}
\[
{g_{\mu \nu} = {\eta}_{ab}~e^{a}_{\mu}~e^{b}_{\nu} \hspace*{2cm} ({\eta}_{ab} =
{\mbox{diag}}(-1, +1, +1)),}
\]
{we see that the subsets of $M$ at which $e^{0}_{\mu}$ 
(the `timelike' vector in the
triad)
vanishes will be sets at which the metric $g_{\mu \nu}$ fails to be
invertible. A
partial analysis of the causal structure of such degenerate spacetimes was
given
above (although only for the case where the singular set is 
a collection of points). The point is, only by allowing such singular behaviour can we complete the
identification of pure gravity as a gauge theory in $(2 + 1)$ dimensions [19].

Having decided on a kinematical framework, the next step is to write down the
Einstein-Hilbert action}
\begin{equation}
{{\cal S}_{EH}[M] = {\int_{M}}e^{a} ~{\wedge}~ (dw_{a} ~+~
{\frac{1}{2}}~{\varepsilon}_{abc}\;w^{b} ~{\wedge}~ w^{c}) = {\int_{M}}e^{a}
{}~{\wedge}~
R_{a}}
\end{equation}
{Thus, the partition function, $Z(M)$, associated with any closed
three-manifold $M$
(with or without boundary) is given abstractly by the path integral}
\begin{equation}
{Z(M) = {\int}{\cal D}w~{\cal D}e~e^{i{\int_{M}}~e^{a}R_{a}}}
\end{equation}
{Integrating over the space of triad fields $e$, one obtains [19]}
\begin{equation}
{Z(M) = {\int}{\cal D}w~{\delta}(R)}
\end{equation}
{where, as in equation (7), $R$ is the ${\frak{so}}~(2, 1)$-valued
curvature 2-form.
Of crucial importance are the boundary conditions one imposes on the connection
1-forms which appear in the sum (8). That is, for a closed three-manifold $M$
with
boundary ${\partial}M$, the data on ${\partial}M$ corresponds to `asymptotic
states'.
The most common approach is to impose the condition that the sum in equation
(8) is
only over flat connections which induce spacelike data on ${\partial}M$ (recall
that
the sum in (8) is over flat connections, since the field equations induced by
equation
(6) imply that the ${\delta}$-function in equation (8) has support only on
those
connections for which $R(w) = 0$). The condition that ${\partial}M$ is
spacelike
relative to the data $(e, w)$ is tantamount to the assumption that $w$ induces
an
$SO(2)$ (or $U(1)$) valued connection 1-form $w |_{{\partial}M}$ on
${\partial}M$.
Amano and Higuchi [21] have shown that imposing these boundary conditions in
the sum
(8) leads to a selection rule which governs when the amplitude for a given
topology-changing process is non-vanishing, i.e. $Z(M) ~{\not=}~ 0$. One of the
principal aims of this paper is to allow for $w$ to have arbitrary boundary
conditions, so that the boundary conditions are classified purely by the
homotopy type
of the data $(e, w)$ on ${\partial}M$. As we shall see, this framework leads
naturally
to a theory which describes processes involving the creation and annihilation
of
gravitational kinks.

At any rate, in order to complete the calculation of the sum in (8), one gauge
fixes
and finds that the total amplitude, $Z(M)$, is equal to the sum over all flat
connections $w$ on $M$ which have the appropriate boundary conditions on
${\partial}M$, and that the contribution to the sum at each such flat
connection $w$
is equal to the ratio of determinants}
\begin{equation}
{{\frac{({\det}\;{\Delta}(w))^{2}}{|{\det}\;L{\_}(w)|}}}
\end{equation}
{where ${\Delta}(w)$ is a certain Laplacian constructed using $w$, and
$L{\_}(w)$ is a
`twisted' Dirac operator, also associated with $w$ (see [19]). The ratio in (9)
is
known as the Ray-Singer torsion, and it is a topological invariant, i.e. it
does not
depend on the choice of metric made in gauge fixing. Thus, the total amplitude
$Z(M)$
is also manifestly a topological invariant. Therefore this toy model is a
simple
example of a `topological quantum field theory'. The aim of this paper is to
develop the
mathematical structure so that the `asymptotic states' in the theory can have
extra,
purely topological, degrees of freedom, and to interpret these degrees of
freedom in
terms of gravitational skyrmions.  We note that this construction was first
suggested in [18].}\\
\vspace*{0.6cm}

{\noindent \bf 4. Mathematical Framework}\\

{As we have seen, when studying the structure of a spacetime $(M, g_{L})$,
there are two principal
aspects of `topology' which we will generically encounter: The topology of the
{\it manifold} $M$ (or submanifolds of $M$) and the topology of the {\it
metric}
$g_{L}$. Two manifolds are topologically equivalent iff they are homeomorphic.
Two
metrics are topologically equivalent iff they are homotopic. The problem of
classifying metrics up to homotopy type in $(2 + 1)$ dimensions has been
investigated
elsewhere [27]. Explicitly, in [27] the homotopy classification problem was
solved on
spacetimes of the form ${\Sigma} ~{\times}~ {\Bbb R}$, where ${\Sigma}$ is any
Riemann
surface. It turns out that a (non-singular) metric $g$ on ${\Sigma} ~{\times}~
{\Bbb
R}$ is homotopic to another (non-singular) metric $g^{\prime}$ on ${\Sigma}
{}~{\times}~
{\Bbb R}$ if and only if the kink number of $g$ with respect to ${\Sigma}$ is
equal to
the kink number of $g^{\prime}$ with respect to ${\Sigma}$. This is to be
expected,
since after sll the kink number of a Lorentz metric $g$ with respect to a
two-surface
${\Sigma}$ is precisely the degree of the map, from ${\Sigma}$ to $S^{2}$,
defined by
any timelike vector field associated with $g$. Since we are not working in the
metric formulation of general relativity, but rather in the connection
formalism, we
must transcribe the above discussion and understand when two {\it triads}, $e$
and
$e^{\prime}$, are homotopic. If $v_{0}$ denotes the vector dual to the
`timelike'
1-form $e_{\mu}^{0}$, and $v^{\prime}_{0}$ is likewise dual to
$(e_{\mu}^{0})^{\prime}$, it
is easy to see that the two triads are in the same homotopy class $(e ~{\sim}~
e^{\prime})$ if and only if  ${\mbox{kink}}({\Sigma}; v_{0}) =
{\mbox{kink}}({\Sigma};
v_{0}^{\prime})$, where again we assume that the spacetime has the form
${\Sigma} ~{\times}~
{\Bbb R}$. Henceforth, we shall simply write `${\mbox{kink}}({\Sigma}; e)$' to
denote
the integer classifying the homotopy class of $e$ with respect to ${\Sigma}$
(see [25],
[26] and [27] for the further properties of kinks).

What does this have to do with gravitational skyrmioms?

Well, recall that a skyrmion [29], in any field theory, is simply an extended
structure corresponding to a field configuration with non-trivial topology. It
therefore makes sense to use the terms `kink' and `skyrmion' interchangeably
[25].

Suppose then, that $M$ is a three-manifold with boundary ${\partial}M$ equal to
the
disjoint union of a finite number of closed, orientable two-manifolds:}
\[
{{\partial}M ~{\cong}~ {\Sigma}_{1} ~{\cup}~ {\Sigma}_{2} ~{\cup}~ {\Sigma}_{3}
{}~{\cup}~ ... ~{\cup}~ {\Sigma}_{n}}
\]
{In general, it may be the case that there exists flat $ISO(2, 1)$ data $(e,
w)$ on
$M$ with non-trivial homotopy with respect to ${\partial}M$, i.e. for some
component(s) ${\Sigma}_{i}$ of ${\partial}M$,}
\[
{{\mbox{kink}}({\Sigma}_{i}; e) ~{\neq}~ 0}
\]
{If we take any collar neighbourhood ${\Sigma}_{i} ~{\times}~ (0, 1)$ on which
$e$ is
non-singular, we thus see that ${\mbox{kink}}({\Sigma}_{i}; e)$ classifies the
homotopy type of the data on ${\Sigma}_{i} ~{\times}~ (0, 1)$. If we interpret
a given
collar neighbourhood ${\Sigma}_{i} ~{\times}~ (0, 1)$ as an `incoming' (or
outgoing)
asymptotic state, we see that there are thus two pieces of topological
information
associated to any asymptotic state: The topology of the boundary component
${\Sigma}_{i}$ and the topology of the flat bundle (associated with the flat
data
$(e, w)$) relative to ${\Sigma}_{i}$. Let us adopt the notation}
\[
{{\mbox{kink}}({\Sigma}_{i}; e) = k_{i}}
\]
{The we shall denote an `incoming' state using the ket notation:}
\[
{|~({\Sigma}_{\mbox{in}}, k_{\mbox{in}})>}
\]
{Likewise, we use bra notation for `outgoing' states':}
\[
{<({\Sigma}_{\mbox{out}}, k_{\mbox{out}})~|}
\]
{We wish to extend the selection rules of [21], which dealt with the special
case
$k_{\mbox{in}} = k_{\mbox{out}} = 0$ (referred to as the `spatial sector'). As
we shall see, by allowing for skyrmion
annihilation/production $(k_{\mbox{in}} ~{\neq}~ 0, k_{\mbox{out}} ~{\neq}~
0)$, more
exotic transitions are allowed.}\\
\vspace*{0.6cm}

{\noindent \bf 5. ~Selection Rules and Suppressed Processes}\\

{Let $M$ be a closed three-manifold with boundary equal to the disjoint union
of a
finite collection of closed orientable two-manifolds ${\Sigma}_{i}$, as above,
and let
$k_{i}$ denote the kink number (with respect to a given ${\Sigma}_{i}$) of some
flat
$ISO(2, 1)$ data $(e, w)$ on $M$. We are concerned with understanding when flat
data
$(e, w)$ inducing the given boundary conditions {\it exists}, since given the
existence of such a connection, it necessarily follows that the amplitude (8)
must be
non-vanishing. Likewise, if such flat data does not exist, then the amplitude
for
the process must vanish. Indeed, one readily sees that the three-manifold $M$
mediating the transition is irrelevant, and that the real problem is thus to
classify
which collections of data ${\{}({\Sigma}_{i}, k_{i}){\}}$ correspond to
processes with
non-vanishing amplitude.

Thus, in analogy with the construction in [21], let $P_{i}$ denote an $SO(2,
1)$ bundle
over ${\Sigma}_{i}$ corresponding to the data with homotopy $k_{i}$. It
follows, by
standard formulae, that the Euler class of the $SO(2, 1)$ bundle over
${\Sigma}_{i}$
depends upon both the topology of ${\Sigma}_{i}$ and the `winding' of the
bundle:}
\begin{equation}
{{\mbox{eul}}(P_{i}) = {\pm}\;({\chi}_{i} - {\mbox{kink}}({\Sigma}_{i}; e))}
\end{equation}
{where ${\chi}_{i} = {\chi}({\Sigma}_{i})$.

On the other hand, the sum of all the Euler classes will still equal an
integral over
the boundary of a two-form corresponding to the kink density}
\begin{equation}
{{\sum_{i = 1}^{n}}~{\mbox{eul}}(P_{i}) = {\pm}{\sum_{i = 1}^{n}}~k_{i}}
\end{equation}
{where the $({\pm})$ in equation (10) is the same choice as in equation (11).
We thus
have proved}
\vspace*{0.2cm}

{\noindent {\bf Theorem.} {\it Let ${\{}({\Sigma}_{i}, k_{i})~|~i=1, ... n{\}}$
be some
collection of asymptotic states. Then}
\[
{<({\Sigma}_{n}, k_{n}), ({\Sigma}_{n - 1}, k_{n - 1}), ... ({\Sigma}_{j + 1},
k_{j +
1})~|~({\Sigma}_{j}, k_{j}), ... ({\Sigma}_{1}, k_{1})> ~{\neq}~ 0}
\]
{if and only if}
\[
{{\sum_{i = 1}^{n}}~{\varepsilon}^{i}({\chi}_{i} - 2k_{i}) = 0}
\]
{for some choice ${\varepsilon}^{i} = {\pm}1$.}
\vspace*{0.2cm}

{\noindent {\it Example 1.} Let ${\Sigma} ~{\cong}~ S^{2}$, $k = 1 =
{\mbox{kink}}({\Sigma}; e)$. Then by the above Theorem, we have}
\[
{<(S^{2}, 1)~|~{\phi}> ~{\neq}~ 0}
\]
{In other words, it is possible to nucleate a single `$S^{2}$ universe' from
nothing,
as long as one creates a kink in the process. One can realise this process by
simply
embedding $S^{2}$ in $(2 + 1)$-Minkowski space (so that $M ~{\cong}~ B^{3}$).}
\vspace*{0.2cm}

{\noindent {\it Example 2.} Let ${\Sigma}$ denote the connected sum of $n$
tori}
\[
{{\mbox{tori}}:~ {\Sigma} ~{\cong}~ ~{\overbrace{T ~\#~ T ~\#~ ... ~\#~ T}^{n
\rm\; times}} .}
\]
{Let $k = -1 + n$. Then again, using the Theorem, we find}
\[
{<({\Sigma}, n - 1)~|~{\phi}> ~{\neq}~ 0}
\]
{Again. this process can be visualised by simply embedding the handlebody
${\Sigma}$
in $(2 + 1)$-Minkowski space.}\\
\vspace*{0.6cm}

{\noindent \bf 6. ~Conclusion and Acknowledgements}\\

{We have extended the work of [21] to a selection rule that governs which types
of
skyrmionic creation/annihilation processes are allowed in the theory. We note
that
such processes are only defined for Lorentzian signature; it seems that one
loses
information in passing to a Euclidean formalism. Presumably this issue should
be
explored further.

The author would like to thank Dr. G.W. Gibbons, Dr. L.J. Alty, Dr. T.J. Foxon,
N. Lambert and S.F. Ross for
useful discussions. Also eternal gratitude and thanks go to Jo Chamblin
(Piglit) for
loving support and help with the preparation of this paper. This work was
supported by
NSF Graduate Fellowship No. RCD-9255644.}\\
\vspace*{0.6cm}

{\noindent \bf References}\\

{\noindent [1] Geroch R.P., {\it Topology in General Relativity}, J. Math.
Phys., {\bf
8}, No. 4, 1967}\\
\\
{\noindent [2] Horowitz G.T., {\it Topology Change in Classical and Quantum
Gravity},
Class. Quant. Grav., {\bf 8}, 1991}\\
\\
{\noindent [3] Chamblin A. and Penrose R., {\it Kinking and Causality}, Twistor
Newsletter, {\bf 34}, pgs. 13--18, 1992}\\
\\
{\noindent [4] Bass R.W. and Witten L., Rev. Mod. Phys., {\bf 29}, pg. 452,
1957}\\
\\
{\noindent [5] Steenrod N., {\it Topology of Fibre Bundles}, Princeton Univ.
Press,
1951}\\
\\
{\noindent [6] Hector G. and Hirsch U., {\it Introduction to the Geometry of
Foliations}, Parts A and B, Aspects of Mathematics, Friedr. Vieweg {\&} Sohn
Verlagsgesellschaft mbH, Braunsschweig, 1981}\\
\\
{\noindent [7] Sorkin R., {\it Consequences of Spacetime Topology}, Proc. of
the Third
Canadian Conf. on Gen. Rel. and Rel. Astrophys.}\\
\\
{\noindent [8] Hawking S.W., {\it Spacetime Foam}, Nucl. Phys. B, {\bf 144},
pg. 349,
1978}\\
\\
{\noindent [9] Ashtekar A., {\it New Variables for Classical and Quantum
Gravity},
Phys. Rev. Lett., {\bf 57}, pg. 2244, 1986}\\
\\
{\noindent [10] Witten E., {\it $2 ~+~ 1$ Dimensional Gravity as an Exactly
Soluble
System}, Nucl. Phys. B, {\bf 311}, pg. 46, 1988}\\
\\
{\noindent [11] Hawking S.W. amd Ellis G., {\it The Large-Scale Structure of
Spacetime}, CUP, 1973}\\
\\
{\noindent [12] Finkelstein D. and Misner C., {\it Some New Conservation Laws},
Annals
of Physics, {\bf 6}, pgs. 230--243, 1959}\\
\\
{\noindent [13] Gibbons G.W. and Hawking S.W., {\it Kinks and Topology Change},
Phys.
Rev. Lett., {\bf 69}, No. 12, 1992}\\
\\
{\noindent [14] Chamblin A., Gibbons G.W. and Steif, A.R., {\it Kinks and Time
Machines}, Phys. Rev. D, {\bf 50}, pgs. 2353--2355, 1994}\\
\\
{\noindent [15] Borde A., {\it Topology Change in Classical General
Relativity}, gr-qc
preprint No. 9406053}\\
\\
{\noindent [16] Yodzis, P., {\it Lorentz Cobordism}, Comm. Math. Phys., {\bf
26},
p.39, 1994}\\
\\
{\noindent [17] Milnor, J., {\it Morse theory}, Princeton Univ. Press,
Princeton, NJ, 1951}\\
\\
{\noindent [18] Gibbons, G.W., {\it Topology Change: Kinks and Spinors},
Proc. of the International Workshop of Theoretical Physics, 6th Session:
``String Quantum Gravity and Physics at the Planck Energy Scale'', Erice,
Sicily, 21-28 June (1992)}\\
\\
{\noindent [19] Witten E., {\it Topology Changing Amplitudes in
$2 + 1$-Dimensional
Gravity}, Nucl. Phys. {\bf B 323}, 113--140, 1990}\\
\\
{\noindent [20] Carlip S. and Cosgrove R., {\it Topology Change in $(2 +
1)$-Dimensional Gravity}, J. Math. Phys. {\bf 35}, No. 10, 5477--5493, 1994}\\
\\
{\noindent [21] Amano K. and Higuchi S., {\it Topology Change in $ISO(2, 1)$
Chern-Simons Gravity}, Nucl. Phys. {\bf B 377}, 218, 1992}\\
\\
{\noindent [22] Carlip S. and de Alwis S., {\it Wormholes  in $2 + 1$
Dimensions}, Nucl. Phys. {\bf B 337}, 681, 1990}\\
\\
{\noindent [23] Ashtekar A., Husain V., Rovelli C., Samuel J., and Smolin L.,
{\it $2 + 1$ Quantum Gravity as a Toy Model for the $3 + 1$ Theory}, Class.
Quant.
Grav. {\bf 6}, L185--L193, 1989}\\
\\
{\noindent [24] Carlip S., {\it Exact Quantum Scattering in $2 + 1$-Dimensional
Gravity}, Nucl. Phys. {\bf B 324}, 106, 1989}\\
\\
{\noindent [25] Finkelstein D., {\it Kinks}, J. Math. Phys. {\bf 7}, No. 7,
1218,
1966}\\
\\
{\noindent [26] Williams J.G. and Zvengrowski P., {\it Spin in Kink-Type Field
Theories}, Int. J. Theor. Phys. {\bf 16}, No. 10, 755--761, 1977}\\
\\
{\noindent [27] Williams J.G. and Zvengrowski P., {\it Kink Metrics in $(2 +
1)$-Dimensional Spacetime}, J. Math. Phys. {\bf 33}, No. 1, 256, 1992}\\
\\
{\noindent [28] Witten E., {\it Quantum Field Theory and the Jones Polynomial},
Comm. Math. Phys. {\bf 121}, 351--399, 1989}\\
\\
{\noindent [29] Skyrme T., {\it The Origins of Skyrmions}, Int. J. Mod. Phys.
{\bf A 3}, 2745--2751, 1988}\\
\\
{\noindent [30] Alty L.J., {\it The Generalised Gauss-Bonnet-Chern Theorem},
DAMTP preprint No. R94/ 1994}\\
\\

\end{document}